\documentclass{elsart}
\usepackage{graphicx}

\begin{document}
\begin{frontmatter}

\title{Negative-coupling resonances in pump-coupled lasers}

\author[carr]{T.W. Carr \corauthref{cor}},
\corauth[cor]{Corresponding author.}
\ead{tcarr@smu.edu}
\author[carr]{M.L. Taylor}
\address[carr]{Department of Mathematics\\
Southern Methodist University\\
Dallas, TX 75275-0156\\
tel: 214-768-3460 fax: 214-768-2355}
\author[sch]{I.B. Schwartz \thanksref{schthanks}}
\address[sch]{Nonlinear Dynamical Systems Section, Code 6792\\
Plasma Physics Division\\
Naval Research Laboratory \\
Washington, DC 20375\\
tel: 202-404-8359 fax: 202-767-0631}
\ead{schwartz@nlschaos.nrl.navy.mil}
\thanks[schthanks]{I.B.S. acknowledges the support of the Office of Naval Research.}

\begin{abstract}
We consider coupled lasers, where the intensity deviations from the steady
state, modulate the pump of the other lasers. Most of our results are for two
lasers where the coupling constants are of opposite sign. This leads to a Hopf
bifurcation to periodic output for weak coupling. As the magnitude of the coupling
constants is increased (negatively) we observe novel amplitude effects such as a
weak coupling resonance peak and, strong coupling subharmonic resonances and chaos.
In the weak coupling regime the output is predicted by a set of slow evolution amplitude
equations. Pulsating solutions in the strong coupling limit are described by
discrete map derived from the original model.
\end{abstract}

\end{frontmatter}
\section{Introduction}

In recent work, we presented experimental and simulation results
for two coupled lasers \cite{KiRoArCaSc05} with asymmetric coupling.
That is, the coupling strength from laser-1 to
laser-2 was kept fixed, while the coupling strength from laser-2 to laser-1
was used as a control parameter. In this paper, we present a
more theoretical exploration of the dynamics that result from this
coupling configuration.
Each laser is tuned such that it emits a stable constant light output.
Light-intensity deviations from the steady state are converted to an
electronic signal that controls the pump strength of the other laser.
Our work in \cite{KiRoArCaSc05} considered asymmetric coupling and,
more specifically, the effect of delaying the coupling signal from one
laser to another. The present paper is our first theoretical analysis
of two pump-coupled lasers with asymmetric coupling, but without delay
(analysis of the case with delay will be presented in a future manuscript).
However, we do invert one of the electronic coupling signals such that
the effective coupling constant is negative; for harmonic signals
with delay coupling by half the period, both lead to the same phase
shift. Thus, the present paper serves as a prelude to a future study
of two pump-coupled lasers with same-sign delay coupling.

For very weak coupling, both lasers remain at steady state. As
the coupling is increased, but still small, there is a Hopf bifurcation
to oscillatory output. In the weak-coupling regime we also observe and
describe a resonance peak where the amplitude of both lasers becomes
large over a small interval of the coupling parameter; to our knowledge,
this phenomenon has not previously been reported. For strong coupling,
the oscillations of one laser remain small and nearly harmonic
while the other laser exhibits pulsating output. Period-doubling bifurcations
to chaos and complex subharmonic resonances also exist throughout the
parameter regime. We combine both weakly- and strongly-nonlinear asymptotic
methods to describe the output in the case of strong coupling.

We consider two class-B lasers \cite{ArLiPuTr84,AbMaNa88} modeled by
rate equations as
 \begin{eqnarray}
\frac{dI_{j}}{dt} & = & (D_{j}-1)I_{j},\quad j=1,2 \label{eq:IDmodel}\\
\frac{dD_{j}}{dt} & = & \epsilon_{j}^{2}[A_{j}-(1+I_{j})D_{j}],
\end{eqnarray}
 where $I_j$ is intensity and $D_j$ is the population inversion of each laser. Dimensionless time $t$
is measured with respect to the cavity-decay time $k_{0}$, $t=k_{0}t_r$,
where $t_r$ is real time.
The parameters are
\begin{equation}
\epsilon^{2}=\frac{\gamma_{c}}{k_{0}},\quad A=\frac{\gamma_{\parallel}g}{\gamma_{c}k_{0}}P,
\end{equation}
where $\epsilon^{2}$ is a ratio of the inversion-decay
time, $\gamma_{c}$, to the cavity-decay time, $k_{0}$, and $A$ is proportional
to the pump $P$ (for notational clarity we have suppressed the subscript
$j$ on the parameters in the definitions of $\epsilon$ and $A$).
To facilitate further analysis, we define new variables for the deviations
from the non-zero steady state (CW output) \cite{ScEr94}
$D_{j0}=1$, $I_{j0}=A_{j}-1$ as
\begin{equation}
	I_{j}=I_{j0}(1+y_{j}),\quad D_{j}=1+\epsilon_{j}\sqrt{I_{j0}}x_{j},\quad
	s=\epsilon_{1}\sqrt{I_{10}}t.
\end{equation}
Our goal is to investigate the effects of coupling through the pump with
\begin{equation}
	A_{j}= A_{j0}+I_{j0}\delta_{k}y_{k}.
\end{equation}
Thus, we feed the intensity \textit{deviations} $y_k = (I_k-I_{k0})/I_{k0}$ from the
CW output of laser $k$ to the pump of laser $j$; the strength of the coupling is
controlled by $\delta_{k}$. The pump coupling scheme allows for easy electronic
control of the feedback signal.

Finally, we assume that the decay constants of
the two lasers are related by
$\epsilon_{2}=\epsilon_{1}\frac{\sqrt{I_{10}}}{\sqrt{I_{20}}}(1+\epsilon_{1}\alpha).$
The new rescaled equations are
\begin{eqnarray}
	\frac{dy_{1}}{dt} & = & x_{1}(1+y_{1}),\label{eq:xymodel}\nonumber\\
	\frac{dx_{1}}{dt} & = & -y_{1}-\epsilon x_{1}(a_{1}+by_{1})+\delta_{2}y_{2},\nonumber\\
	\frac{dy_{2}}{dt} & = & \beta x_{2}(1+y_{2}),\nonumber\\
	\frac{dx_{2}}{dt} & = & \beta[-y_{2}-\epsilon\beta x_{2}(a_{2}+by_{2})+\delta_{1}y_{1}],
	\label{e:laserxy}
\end{eqnarray}
where
\begin{equation}
	a_{1}=\frac{1+I_{10}}{\sqrt{I_{10}}},\quad
	a_{2}=\frac{\sqrt{I_{10}}(1+I_{20})}{I_{20}},\quad
	b=\sqrt{I_{10}}\quad\textrm{and}\quad\beta=1+\epsilon\alpha.
\end{equation}
For notational convenience we have let $s\rightarrow t$ and dropped the subscript on
$\epsilon_{1}$($\epsilon_{1}\rightarrow\epsilon).$ We mention that Eq.~(\ref{e:laserxy})
is similar to the coupled laser equations studied by Erneux and Mandel \cite{ErMa95} to
investigate antiphase (splayphase) dynamics in lasers. However, antiphase dynamics require
global coupling that would correspond to the symmetric case of $\delta_2 = \delta_1$
in our model.

A main point of interest in the study of coupled oscillators in general is
their degree of synchronization. This implies a focus on the
phase- and frequency-locking characteristics of the oscillators. Thus, many
investigations focus on coupled-phase oscillators (see \cite{ArErKo90} and \cite{St00} for
reviews and extensive bibliographies).
Consideration of just the phase relationships between the oscillators
is often based upon considering limit-cycle oscillators with \textit{weak coupling}.
In that case, each oscillator's amplitude is fixed to that of the limit cycle and only the
phase remains a dynamical variable. However,
limit-cycle oscillators with \textit{strong coupling} can exhibit amplitude instabilities
leading to amplitude death and other novel phenomena \cite{ArErKo90}.

Class-B lasers, which include such common lasers as semiconductor, YAG,
and CO$_2$ lasers, are not limit cycle oscillators, but, rather, are
perturbed conservative systems \cite{ErBaMa87}. (The underlying form of the perturbed
conservative system, Eqs.~(\ref{e:laserxy}) with $\epsilon =0$, has also been
used in population dynamics models \cite{ScSm83}.) Thus, the amplitude
is not fixed by a limit cycle and remains an important dynamical variable.
This has been demonstrated in laser systems coupled by
mutual injection \cite{LiMaEr90} and overlapping evanescent fields
\cite{FaCoroLe93}, or by multimode lasers with coupled modes \cite{WiBrJaRo90,VlViMa99}
to name just a few.
Under certain conditions, phase-only equations can be derived that describe
the behavior of the coupled laser systems \cite{FaCoroLe93,Wi92,HoGaErKo97}.
However, in general, the amplitude cannot be adiabatically removed
and amplitude instabilities can dominate the observed dynamics.
In is interesting to note, however, that a time-dependent phase
is sometimes the drive leading for the laser's observed amplitude instability
\cite{ThMoRoCaLiEr97}.

The coupled laser equations, Eqs.~(\ref{e:laserxy}), have all real coefficients.
If the lasers were coupled directly through their electric fields (referred
to as ``coherent coupling"), such as in evanescent or injection coupling, then
there would be a complex detuning parameter or coupling coefficient.
In Eqs.~(\ref{e:laserxy}) the lasers are coupled through their real intensities
(referred to as ``incoherent coupling") such that the differences in the laser's
optical frequency do not affect the systems dynamics.

In the next section, we give an overview of the laser system's behavior
as the coupling is increased. We begin with the linear-stability analysis
of the CW steady state $(x,y)=(0,0)$ and find that there are two possible
Hopf bifurcations to oscillatory output, one for large $O(1)$ coupling,
and one for small $O(\epsilon)$ coupling; we focus the rest of our
analysis on the latter and continue our overview by presenting results
of numerical simulations over the full range of coupling strengths.
In Sec.~\ref{s:harmonic} we analyze the oscillatory solutions for
weak coupling. We also show how the results in this parameter regime
extend to the case of three or more lasers. In Sec.~\ref{s:strong} we
consider large coupling and combine
the method of multiple scales and matched asymptotics \cite{KeCo96}
to derive a map that describes the coexisting small- and large-amplitude
solutions. Finally, in Sec.~\ref{e:discuss} we discuss and summarize our results.

\section{Bifurcations for negative coupling}

In the new variables, the CW state is given by $x=y=0$.  The linear stability of
the CW state is governed by the characteristic equation
\begin{equation}
	[\lambda(\lambda+\epsilon a_{1})+1][\lambda(\lambda+
	\epsilon a_{2}\beta^{2})+\beta^{2}]-\beta^{2}\delta_{1}\delta_{2}=0.
	\label{e:charequ}
\end{equation}
If both $\delta_{j}=0$, as expected we find that each laser is a damped oscillator.
For $\delta_{j}\ne0$ we study Eq.~(\ref{e:charequ}) for small $\epsilon\ll 1$.
Keeping $\delta_{1}$ as a fixed parameter and varying $\delta_{2}$ we find that
there are two Hopf bifurcations. If $\delta_{2} <0 $ and $|\delta_2|$ increases,
then the condition for a Hopf bifurcation is
\begin{equation}
	\delta_{1}\delta_{2}+\epsilon^{2}[a_{1}a_{2}+
	4\alpha^{2}\frac{a_{1}a_{2}}{(a_{1}+a_{2})^{2}}]+O(\epsilon^3)=0.
	\label{eq:ls-hopf-neg}
\end{equation}
If $\delta_{2}>0$ and $|\delta_2|$ increases the Hopf condition is
\begin{equation}
	\delta_{1}\delta_{2}=1 +O(\epsilon^2).
	\label{eq:ls-hopf-pos}
\end{equation}
The second condition, Eq.~(\ref{eq:ls-hopf-pos}), indicates that a Hopf bifurcation
occurs when there is strong coupling between the lasers, $\delta_1\delta_2 = O(1)$. We
are interested in the Hopf bifurcation that occurs for weak coupling that is described by
the first condition, Eq.~(\ref{eq:ls-hopf-neg}) (this is the relevent case when
the problem is extended to include delayed coupling). In this case,
$\delta_{2H}=O(-\epsilon^{2}/\delta_{1}) < 0$, that is, the coupling from
laser-2 to laser-1 is negative.

\begin{figure}
\includegraphics{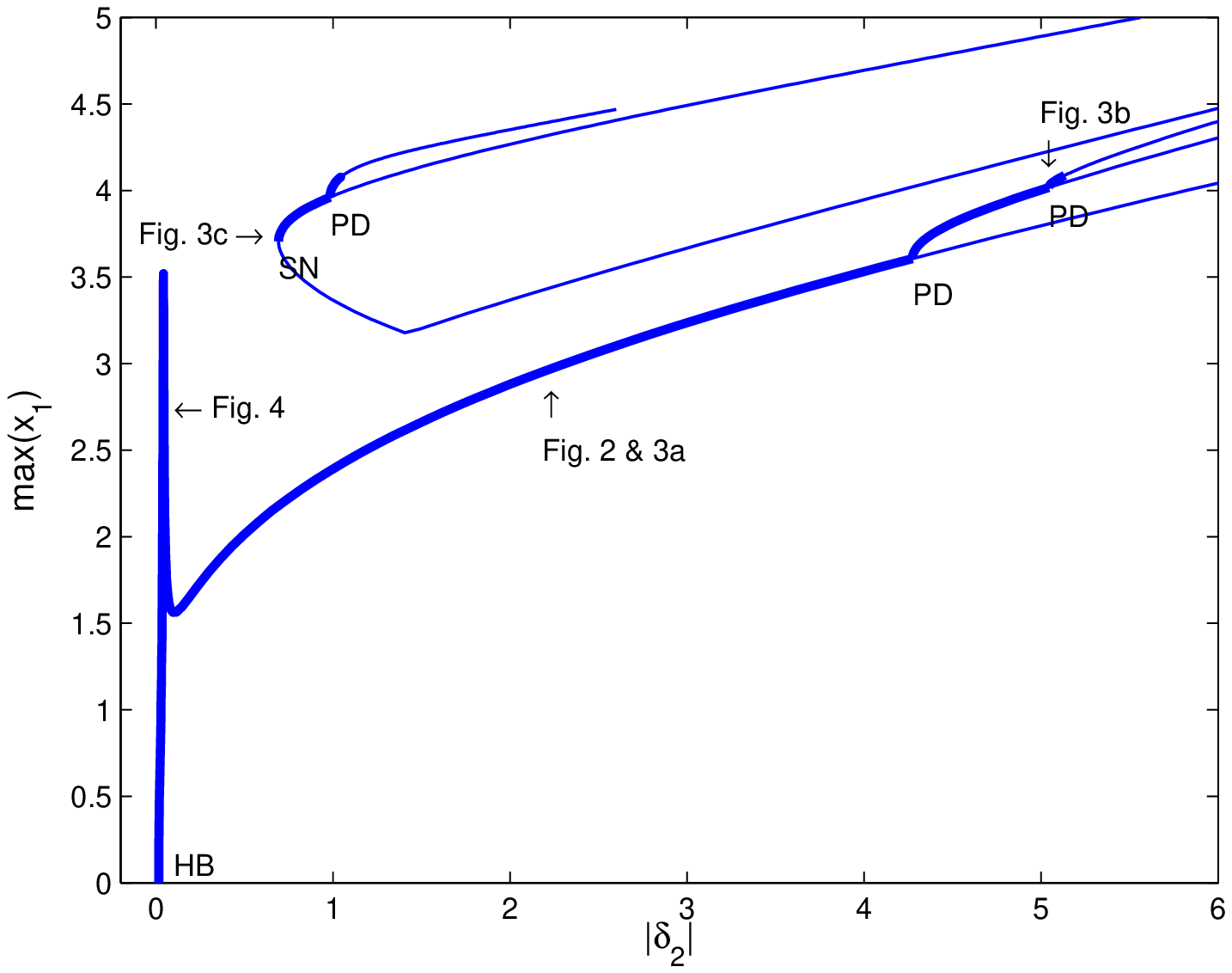}
\caption{Numerical bifurcation diagram (AUTO) \cite{Auto2000}
using $\delta_{2}$ as the bifurcation
parameter. Note, the coupling is \textit{negative} so $\delta_2<0$;
as we increase $|\delta_2|$, the coupling is \textit{increasingly negative.}
Thick (thin) lines indicate that
the periodic solutions are stable (unstable). There is an initial Hopf bifurcation
(HB) from the steady state to periodic solutions.
As $|\delta_2|$ is increased the initial periodic
orbit exhibits a period-doubling (PD) sequence of bifurcations to chaos; we
show only the first two PD bifurcations. Simultaneously, subharmonic periodic
solutions appear through saddle-node (SN) bifurcations and will also period
double. For these parameter values, $|\delta_{2S}| > |\delta_{2H}|$ so that the
small-coupling feedback resonance peak is exhibited. (Fixed parameters are
$\epsilon=0.001$, $b_{1}=b_{2}=1$, $a_{1}=a_{2}=25$ and $\delta_{1}=0.04$.)}
\label{f:bifdiafull}
\end{figure}

In Fig.~\ref{f:bifdiafull} we show the amplitude of the periodic solutions
that emerge from the Hopf bifurcation point of Eq.~(\ref{eq:ls-hopf-neg}).
As the magnitude of the coupling constant $(|\delta_2|,\;\;\delta_2<0)$ is increased,
the Hopf bifurcation leads to small-amplitude periodic solutions. However, for small coupling
there is a strong resonance effect where the amplitudes become $O(1)$. In Fig.~\ref{f:bifdiafull}
this appears as a narrow spike in the amplitude. We show a close-up of the amplitude
resonance in Fig.~\ref{f:reson} (calculated at different parameter values). Both before and
after, the amplitude is small and nearly harmonic, as would be
expected for weak coupling. However, during
the resonance the amplitude is pulsing.

As $|\delta_2|$ is increased, the coupled system behaves similar
to a periodically modulated laser \cite{ScEr94,CaBiScTr00}. The intensity of laser-1
increases and becomes pulsating (see Fig.~\ref{f:timeplot_bpd}a) because the effective
modulation signal from laser-2 becomes stronger. On the other hand, because $\delta_1$
is fixed and small, laser-2 receives only a weak signal from laser-1 and remains
nearly harmonic (see Fig.~\ref{f:timeplot_bpd}b). For larger coupling,
the periodic solutions exhibit a period-doubling sequence to chaos; the inversion for
both laser-1 and -2 after the first period-doubling bifurcation is shown in
Fig.~\ref{f:timeplot_all}b. We mention also that for different parameter values
the original branch of periodic solutions may remain completely stable and not exhibit
further bifurcations.

Coexisting with the primary branch of periodic solutions are subharmonic resonances that
appear through saddle-node bifurcations. These also exhibit period-doubling bifurcations
for increasing coupling. In Fig.~\ref{f:timeplot_all}c we see that just after the
primary saddle-node bifurcation the periods of the oscillations are in a 2:3 ratio,
with 2 maximum of laser-1 for every 3 of laser-2.

\section{Weak-coupling resonance}
\label{s:harmonic}

\subsection{Two lasers}
\label{s:two}

We now describe the periodic solutions that emerge from the Hopf bifurcation
located by Eq.~(\ref{eq:ls-hopf-neg}). We use the standard method of
multiple time-scales \cite{KeCo96} approach and thus only summarize the results.
From the linear-stability
analysis, we know that solutions decay on an $O(\epsilon)$ time scale. This
suggests that we introduce the slow time $T=\epsilon t$, such that $x=x(t,T)$
(similarly for $y$) and time derivatives become
$\frac{d}{dt}=\frac{\partial}{\partial t}+\gamma^{2}\frac{\partial}{\partial T}$.
We analyze the nonlinear problem using perturbation expansions in powers
of $\epsilon^{1/2}$, e.g.,
$x_{j}(t)=\epsilon^{1/2}x_{j1}(t,T)+\epsilon x_{j2}(t,T)+\ldots$,
Finally, we assume that the coupling constants are small and let
$\delta_{j}=\epsilon d_{j}$.

At the leading order, $O(\epsilon^{1/2})$, we obtain the solutions
\begin{equation}
	y_{j1}(t,T)=A_{j}(T)e^{it}+c.c.,\quad x(t,T)=iA_{j}(T)e^{it}+c.c.,
\end{equation}
which exhibit oscillations with radial frequency 1 on the $t$ time scale.
To find the slow evolution of $A_{j}(T)$ we must continue the analysis
to $O(\epsilon^{3/2})$. Then, to prevent the appearance of unbounded
secular terms, we determine ``solvability conditions'' for
the $A_{j}(T)$ that are given by
\begin{eqnarray}
	\frac{dA_{1}}{dT} & = & -\frac{1}{2}a_{1}A_{1}-\frac{1}{6}i|A_{1}|^{2}A_{1}
	-\frac{1}{2}id_{2}A_{2},\\
	\frac{dA_{2}}{dT} & = & -\frac{1}{2}a_{2}A_{2}-\frac{1}{6}i|A_{2}|^{2}A_{2}
	-\frac{1}{2}id_{1}A_{1}+i\alpha A_{2}.
\end{eqnarray}
To analyze these equation we let $A_{j}(T)=R_{j}(T)e^{i\theta_{j}(T)}$ and
consider the phase difference $\psi=\theta_{2}-\theta_{1}$ to obtain
\begin{eqnarray}
	\frac{dR_{1}}{dT} & = & -\frac{1}{2}a_{1}R_{1}+\frac{1}{2}\sin(\psi)d_{2}R_{2},\\
	\frac{dR_{2}}{dT} & = & -\frac{1}{2}a_{2}R_{2}-\frac{1}{2}\sin(\psi)d_{1}R_{1},\\
	\frac{d\psi}{dT}& = & -\frac{1}{6}(R_{2}^{2}-R_{1}^{2})-\frac{1}{2}\cos\psi(d_{1}\frac{R_{1}}{R_{2}}
	-d_{2}\frac{R_{2}}{R_{1}})+\alpha.
\end{eqnarray}
The leading order, solutions are $t=2\pi$ periodic if the amplitudes and phase are constant
with respect to the $T$ time scale (derivatives with respect to $T$ are zero). This
determines the bifurcation equation for the amplitudes $R_j$ and the phase
difference $\psi$ as
\begin{equation}
	R_{2}^{4}=-9\frac{\Delta_1}{\Delta_{2}^2},\mbox{ where }
	\Delta_{1}=(1+\frac{d_{1}d_{2}}{a_{1}a_{2}})(a_{1}+a_{2})^{2},\mbox{ and }
	\Delta_{2}=1+\frac{a_{2}d_{2}}{a_{1}d_{1}},
	\label{e:bifequ}
\end{equation}
and
\begin{equation}
	R_{1}^{2}=-\frac{a_{2}d_{2}}{a_{1}d_{1}}R_{2}^{2},\;
	\cos^{2}\psi=(1+\frac{a_{1}a_{2}}{d_{1}d_{2}}),
	\label{eq:r1phase}
\end{equation}
where we have set $\alpha=0$ to simplify the discussion.
For $R_{1}$ to be positive in Eq. (\ref{eq:r1phase}), $d_{1}$ and
$d_{2}$ must have opposite signs, while the Hopf bifurcation point
is determined by taking $R_{2}\rightarrow 0$ in Eq.~(\ref{e:bifequ}) to obtain
$\Delta_{1}=0$; both conditions are consistent with the
linear stability results in Eq.~(\ref{eq:ls-hopf-neg}). We define the
value at which the Hopf bifurcation occurs to be
$\delta_{2H}=\epsilon d_{2H}$. For $d_2 > d_{2H}$, Eq.~(\ref{eq:r1phase})
describes a supercritical bifurcation to stable periodic solutions;
this is consistent with the numerical bifurcation diagram in
Fig.~\ref{f:bifdiafull}. Finally, because $d\psi/dT=0$, the
laser oscillations are phase locked with the phase difference described
by Eq. (\ref{eq:r1phase}), and the frequency for $x$ and $y$ is
\begin{equation}
   \omega = 1 - \frac{\epsilon}{2}\sqrt{-\Delta_1}
    \left(\frac{1}{|\Delta_2|} + \frac{a_2}{a_1+a_2}\right) + O(\epsilon^{3/2}).
    \label{e:freq}
\end{equation}

An important result of this paper comes from an examination of $\Delta_2$
in Eq. (\ref{e:bifequ}). Specifically, the bifurcation equation is
singular when $\Delta_{2}=0$ or
\begin{equation}
	d_2 = d_{2S}\equiv -\frac{a_{1}}{a_{2}}d_{1}.
	\label{e:delta2s}
\end{equation}
If $d_{2S} < d_{2H}$, then the singularity occurs before
the Hopf bifurcation when the CW steady-state is still stable. Thus, in this case,
the singularity is not seen and does not affect the amplitude of the
bifurcating periodic solutions. However,
if $d_{2S}>d_{2H}$, then near the bifurcation point the amplitude
of the oscillations becomes very large corresponding to a resonance.
The resonance can be understood as a balance between an effective
negative damping due to the coupling term, and the self damping. That is,
the ratio $d_2/a_1$, which is the relative negative camping to the
self damping in laser-1, is equivalent to $d_1/a2$ (modulus
the negative sign), the relative negative damping to self damping in laser-2.
The net result is that the coupling terms provide an effective negative-damping
that cancels with the lasers self-damping and, hence, a resonance effect.

\begin{figure}
\includegraphics[scale=0.5]{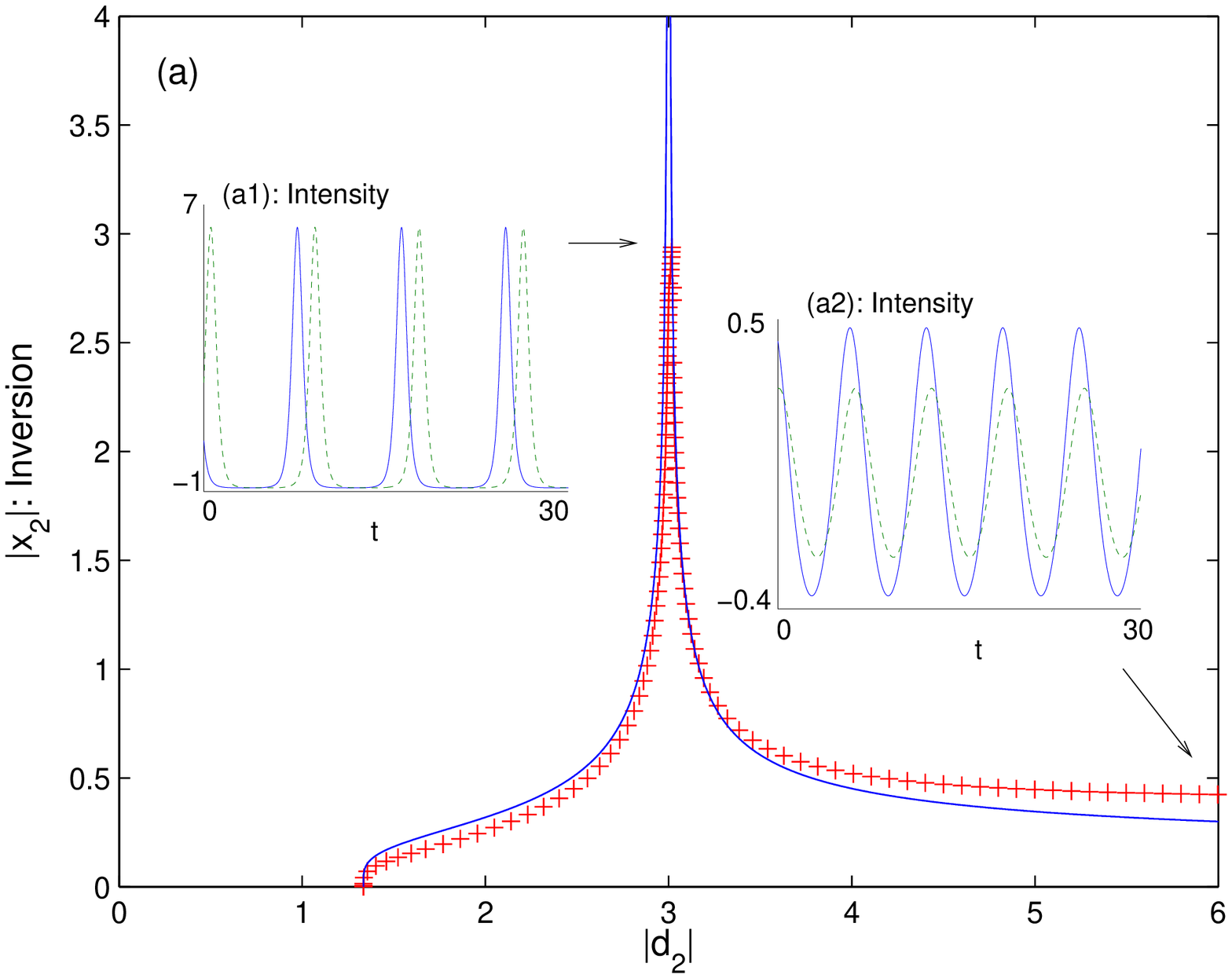}\\
\includegraphics[scale=0.6]{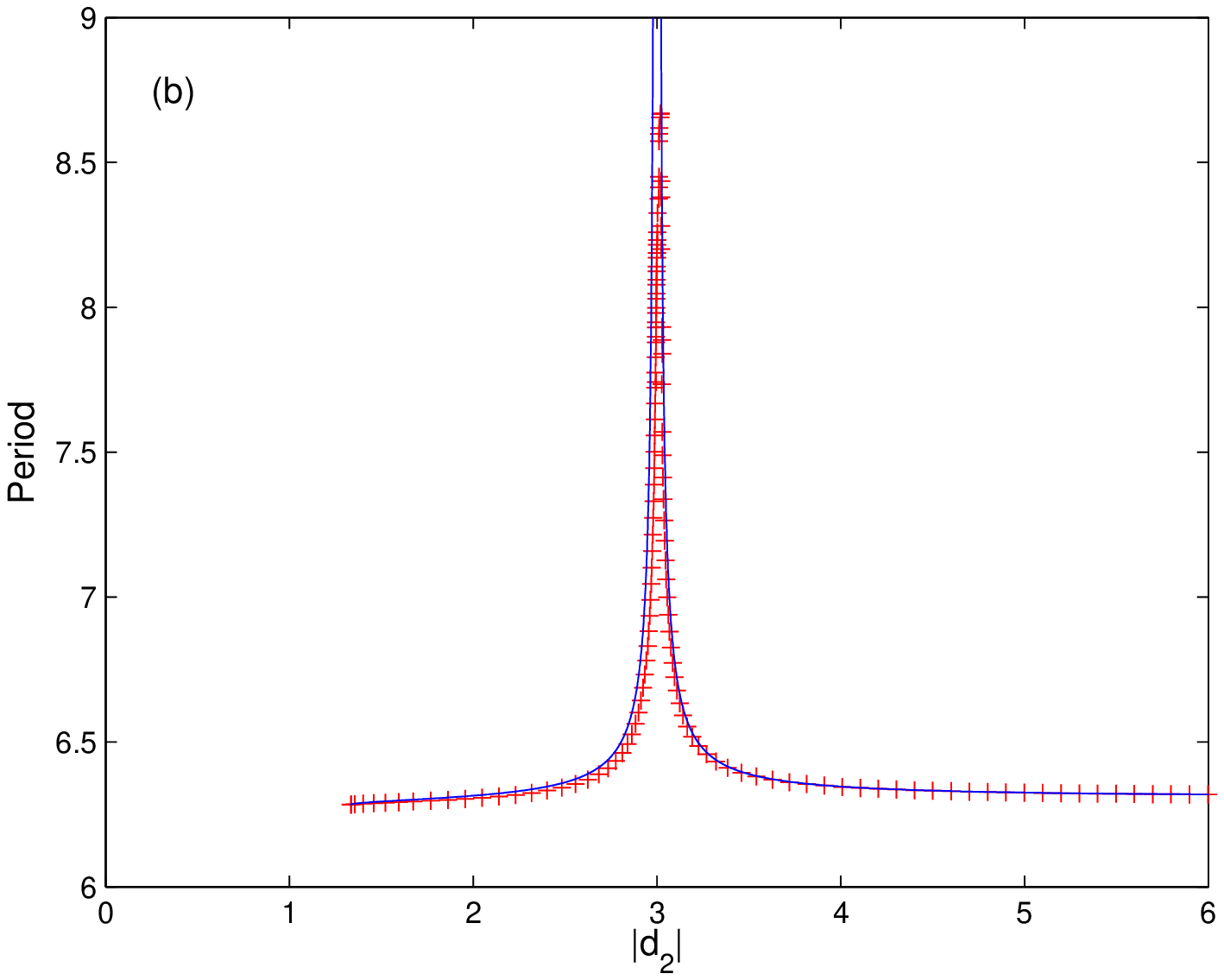}
\caption{(a) Amplitude of \textit{inversion} $x_2$ as a function of $\delta_2 = \epsilon d_2$ after
the Hopf bifurcation; numerical (+) (Auto \cite{Auto2000}), analytical from Eq.~(\ref{e:bifequ})
(solid line). Parameter values are $\epsilon = 0.001$, $a_1 =a_2 = 2$, $d_1=3$  and $\alpha = 0$ so that
$d_{2H} = 4/3$ and $d_{2S} = 3$. In the inset (a1) we show the pulsating \textit{intensity}
near the peak of the resonance ($y_1$ dashed. $y_2$ solid), while in inset (a2),
the \textit{intensity} has returned
to be small-amplitude and nearly harmonic. (b) Period of oscillations as a function
of $\delta_2 = \epsilon d_2$ with analytical result (solid line) from Eq.~\ref{e:freq}.}
\label{f:reson}
\end{figure}

The negative-coupling resonance when $d_2 = d_{2S}$ is demonstrated
in Fig.~\ref{f:reson}a. The solid line is the result of our analytical bifurcation
curve given by Eq.~(\ref{e:bifequ}), while the $+$ are data from numerical
simulation; the analytical and numerical results are in excellent agreement.
In the vicinity of $\delta_2=\delta_{2S}$ the amplitude of the periodic
oscillations become $O(1)$, whereas we would normally expect the amplitude
to remain $O(\epsilon^{1/2})$.

\begin{figure}
\includegraphics[scale=0.75]{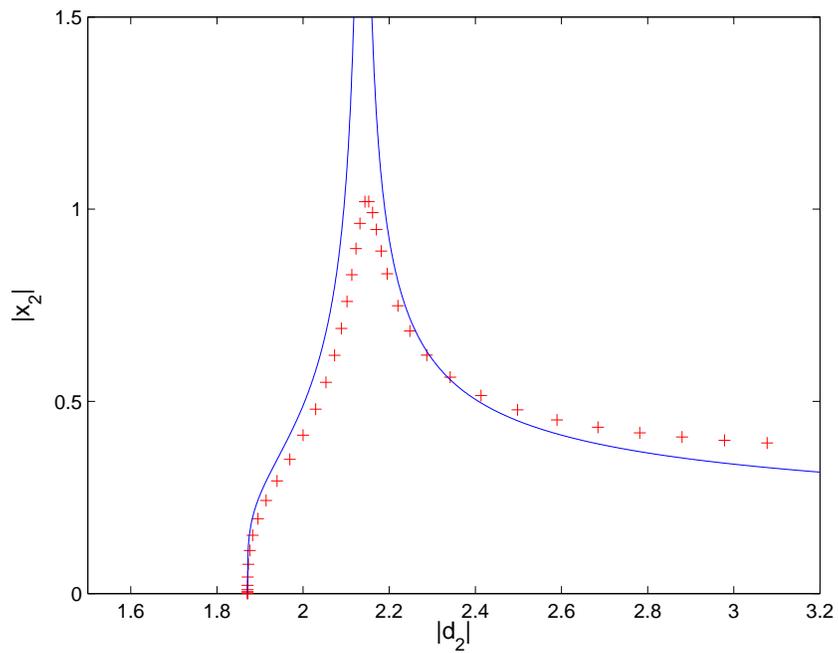}
\caption{Amplitude of $x_2$ as a function of $\delta_2 = \epsilon d_2$ after
the Hopf bifurcation; numerical (+) \cite{Auto2000}, analytical (solid line). The
parameters have been chosen as $a_2=2.9$ and $d_1=3.1$ so that
the singular point $d_{2S}=2.13$ is very near the Hopf bifurcation $d_{2H}=1.87$.}
\label{f:reson_local}
\end{figure}

Comparing Fig.~\ref{f:reson}a and Fig.~\ref{f:reson_local}, we see that the
maximum amplitude, when $\delta_{2}=\delta_{2S}$, depends on the parameters.
However, the bifurcation equation is singular at $\delta_{2S}$ and does
not give a value for the maximum. The bifurcation equation can be improved
by tuning the resonance closer to the Hopf bifurcation point with
$\delta_1 = \epsilon a_2 + O(\epsilon^{3/2})$ and
$\delta_2 = -\epsilon a_1 +O(\epsilon^{3/2})$ and continuing the perturbation
analysis to $O(\epsilon^2)$. Unfortunately, the analysis become algebraically
difficult and we have not pushed through to its conclusion.

During the resonance both lasers become pulsating. Pulsating solutions
are not well described by the weakly-nonlinear analysis of the present section.
In Appendix~\ref{s:pulsations} we consider pulsating lasers and again
locate the resonance peak at $\delta_{2S}$. We discuss this further in
the paper's final discussion section.

\subsection{Three (or more) lasers}
\label{s:three}

The resonance spike can also be found in three or more lasers.
In general, the amplitudes of the periodic solutions near the Hopf bifurcation
are described by coupled Stuart-Landau equations of the form
\begin{equation}
	\frac{dA_{j}}{dT}  =  -\frac{1}{2}a_{j}A_{j}-\frac{1}{6}i|A_{j}|^{2}A_{j}
	-\frac{1}{2}i \sum_{k=1,k \ne j}^{N} d_{jk}A_{k}, \quad j=1\ldots N.
	\label{e:stuartlandau}
\end{equation}
As written, the coupling coefficients are completely general and could be chosen
to give global coupling, $d_{jk}=d$, nearest-neighbor coupling, $d_{j,k} \ne 0$ for only
$k=j+1,\;\; k=j-1$, or any other coupling configuration.
Coupled algebraic equations for the amplitudes $R_j$ can then be found with
the substitution $A_j = R_j \exp(i\theta_j)$. An amplitude resonance
occurs when there is a vanishing denominator in the equation for any one of the
isolated amplitudes $R_j = g(R_k),\;\; k\ne j$. As with two lasers, one of the
coupling constants must be negative to produce the resonance. However, obtaining
an explicit solution for one of the laser amplitudes, even in the case of only three lasers,
is extremely difficult in all but the most trivial cases.

\begin{figure}
\includegraphics{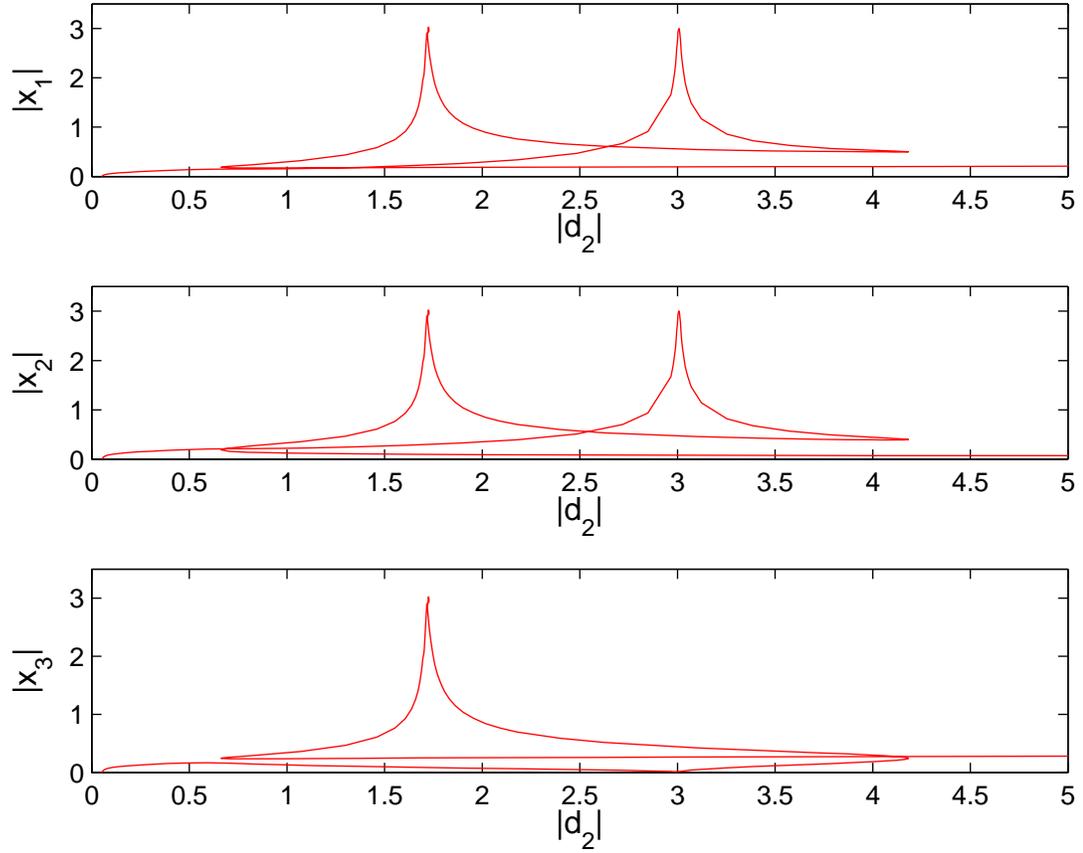}
\caption{Amplitudes for the case of three identical lasers (fixed parameters are
$a_1=a_2=a_3=2$, $b=1$, $\beta_1=\beta_2=1$, $\epsilon= 0.001$.) The fixed-coupling
contants are $\delta_{13}=\delta_{23}=1.3$ and $\delta_{21}=\delta_{31} = 3$.
The coupling of laser-2 into laser-3 is positive with size $\delta_{32}=|\delta_2|$,
while the coupling of laser-2 into laser 1 is negative with $\delta_{12}=\delta_2 <0.$}
\label{f:3lasers}
\end{figure}

\begin{figure}
\includegraphics{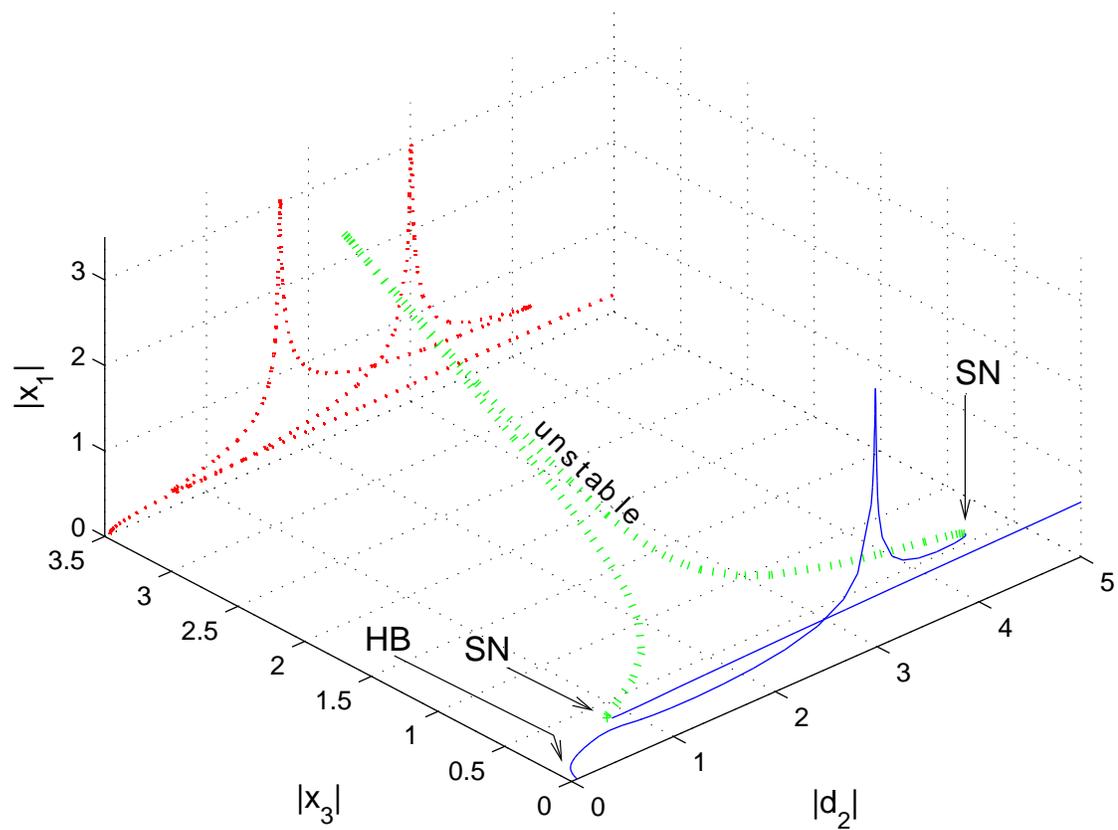}
\caption{Same data as Fig.~\ref{f:3lasers} but with the amplitudes of laser-1 and
laser-3 shown simultaneously. The solid curve shows the stable solutions that
appear after the Hopf bifurcation (HB). There is a stable resonance peak for
$|d_2|\approx 3$. The dashed curve shows
unstable solutions that exist between the two saddle-node (SN) bifurcations.
There is an unstable resonance peak for $|d_2|\approx 1.75$.
Both laser-3 and laser-1 (and laser 2) exhibit large amplitudes during the
unstable resonance, while only laser-1 (and laser-2) has large amplitude
during the stable resonance. The dotted curve is a projection
of the actual data into only the laser-1 plane to compare to laser-1 data in Fig.~\ref{f:3lasers}.
There is bi-stability between the saddle-node (SN) points.}
\label{f:reson3d}
\end{figure}

In contrast, demonstrating
the resonance effect numerically requires only some experimentation and we show
one result in Figs.~\ref{f:3lasers} and \ref{f:reson3d}. In Fig.~\ref{f:3lasers}
we show the amplitude of each laser as a function of one of the coupling parameters.
Specifically, we fix the coupling of laser-3 into laser-1 and laser-2 as
$d_{13}=d_{23}=1.3$ and the coupling of laser-1 into laser-2 and laser-3 as
$d_{21}=d_{31} = 3$. The coupling of laser-2 into laser-3 is positive
with size $d_{32}=|d_2 |$, while the coupling of laser-2 into laser-1 is
negative with $d_{12}=d_2 <0.$ We use $d_2$ as the control parameter.

As $|d_2|$ is increased, both
laser-1 and laser-2 show two resonance peaks, while for laser-3 there is only one.
However, in Fig.~\ref{f:reson3d} we see that the branch of solutions is not
monotonic in $|d_2|$. As the branch of solutions is followed from the Hopf
bifurcation point, the resonance peak for larger $|d_2| \approx 3$ occurs first, but only
for laser-1 and laser-2. As the branch is followed further, it turns at the saddle-node
bifurcation (right most in figure). As $|d_2|$ decreases, all
three lasers exhibit a resonance when $|d_2| \approx 1.75$. The branch turns again
at a saddle-node bifurcation (left most) to then increase without any further resonances.

As it happens, the periodic solutions are unstable on the branch of solutions
with the lower resonance peak $|d_2|\approx 1.75$. Thus, between the two
saddle-node bifurcations there are two stable solutions: the
primary branch originating from the Hopf bifurcation that exhibits a resonance
for laser-1 and laser-2 and terminates at the larger (right) saddle-node bifurcation,
and the small-amplitude branch that appears at the lower (left) saddle-node
bifurcation and continues for $|d_2|>5$. Thus, in the vicinity of the resonance
when $|d_2|\approx 3$ and both stable solutions
coexist, initial conditions will determine whether the large-amplitude resonant
solutions or the small-amplitude solutions are exhibited

Finally, the period of the oscillations shows a sharp peak at each of the amplitude
resonances. The result is analogous to that of the peak in the period for two lasers
as shown in Fig.~\ref{f:reson}b.

\section{Strong coupling}
\label{s:strong}

\begin{figure}
\includegraphics{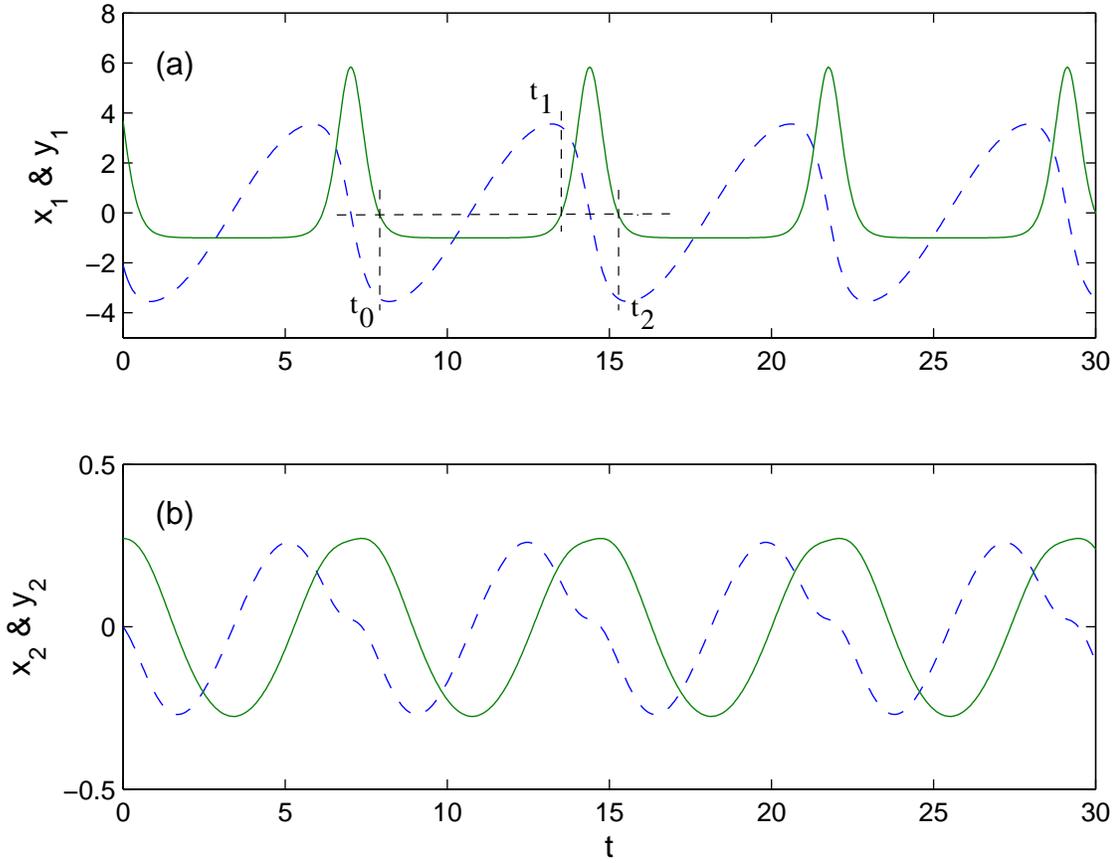}
\caption{For $|\delta_2| = 2.18$ ($\delta_2<0$). (a) Laser-1 has pulsating intensity (solid) and
a triangular-shaped population inversion (dashed) because it is strongly modulated
by laser-2. (b) Laser-2 is nearly harmonic because it receives only weak
coupling from laser-1. In (a), the times marked $t_0$, $t_1$ and $t_2$ define
the outer and inner regimes discussed in Sec.~\ref{s:strong}.}
\label{f:timeplot_bpd}
\end{figure}

\begin{figure}
\includegraphics[scale=0.5]{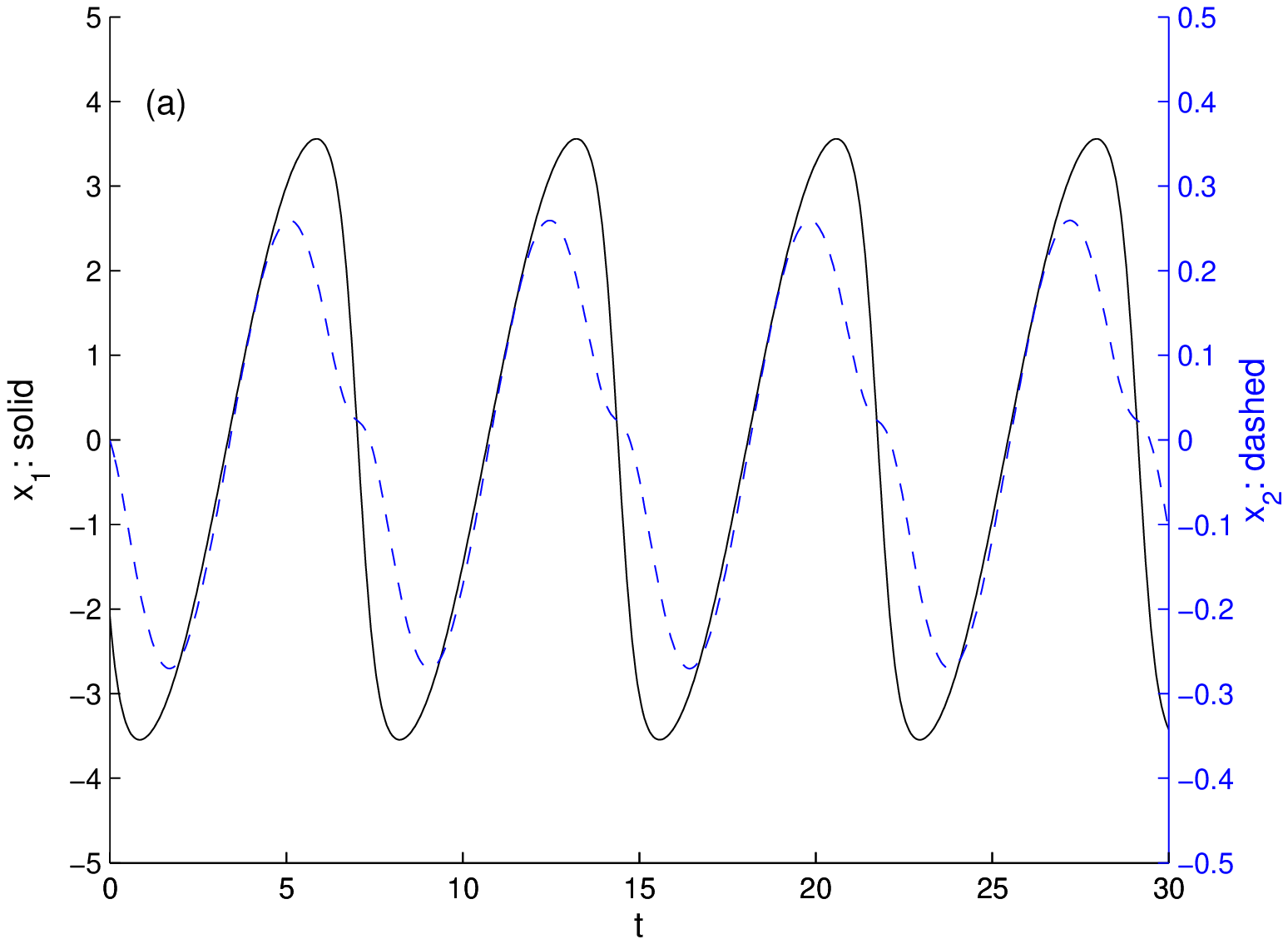}
\includegraphics[scale=0.5]{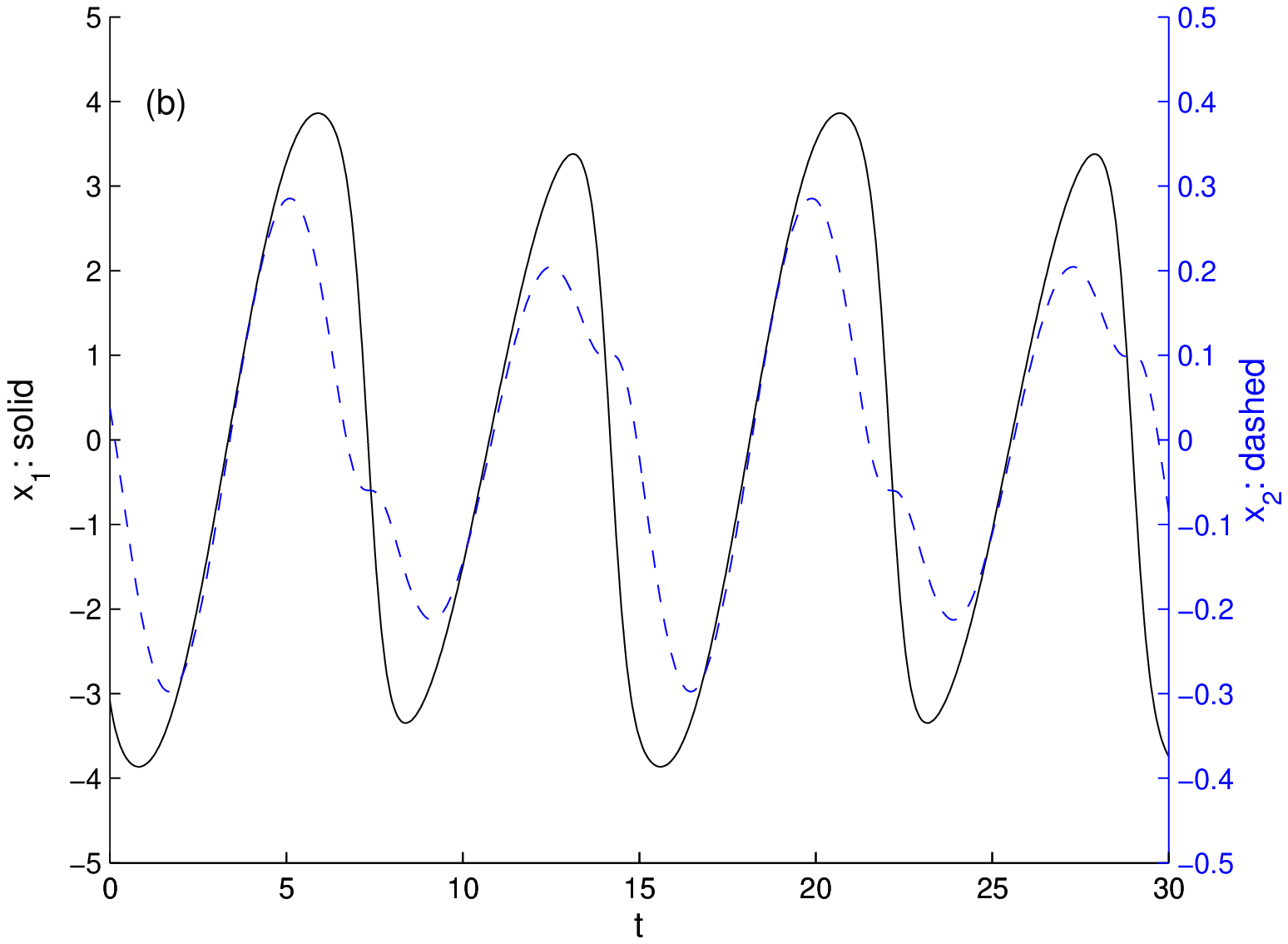}
\includegraphics[scale=0.5]{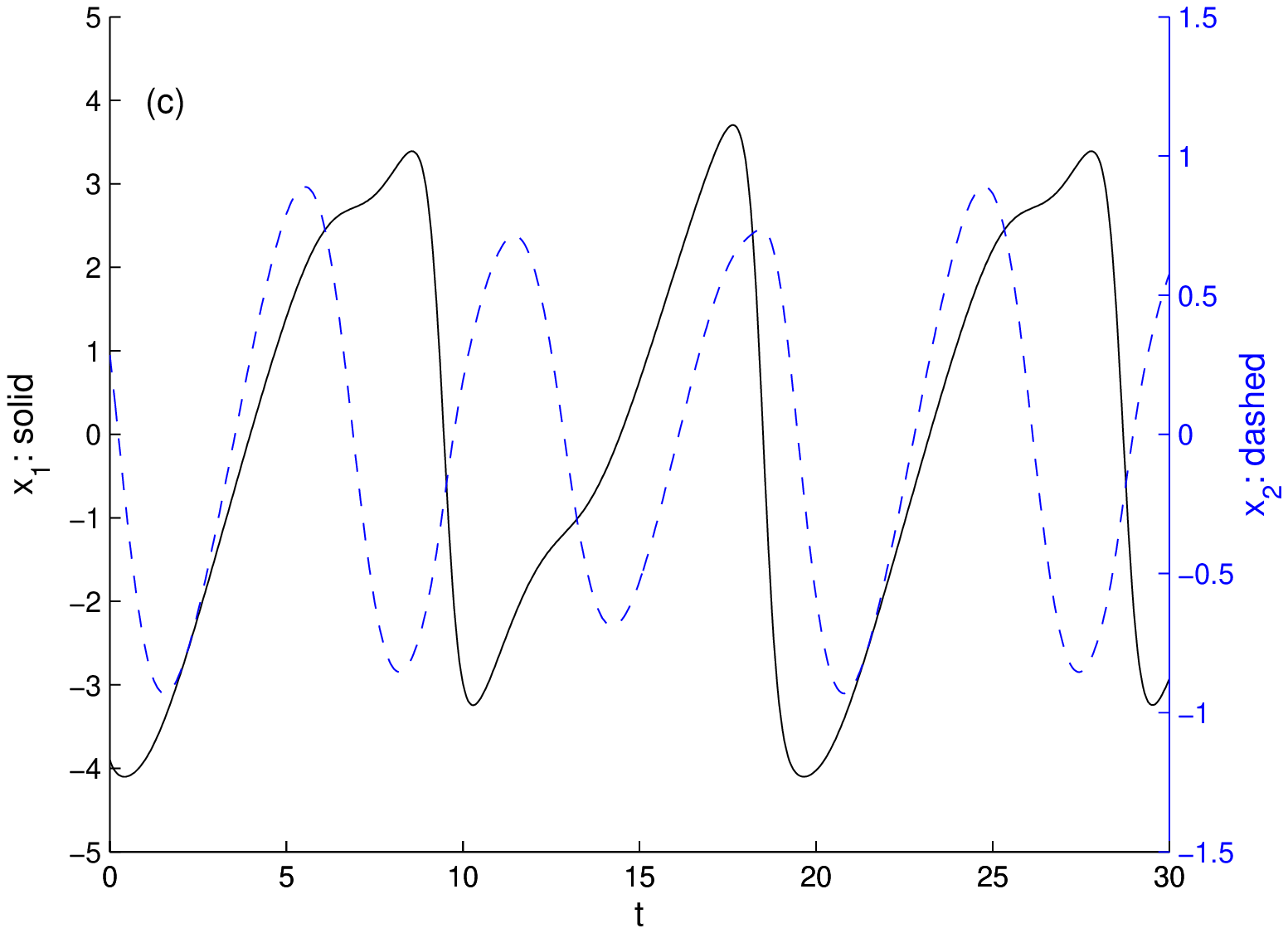}
\caption{Comparison of the amplitude of the population inversion of laser-1 (left axis, solid
curve) and the inversion of laser-2 (right axis, dashed line) for: (a) Before
the first period-doubling bifurcation, $|\delta_2| = 2.18$ (same
as Fig.~\ref{f:timeplot_bpd}). (b) After the period-doubling bifurcation $|\delta_2| = 5.04 $.
(c) Just after the saddle-node bifurcation near the limit point, $|\delta_2| = 0.70 $.
There are three maximums of laser-2 for every two maximum of laser-1.}
\label{f:timeplot_all}
\end{figure}

\begin{figure}
\includegraphics{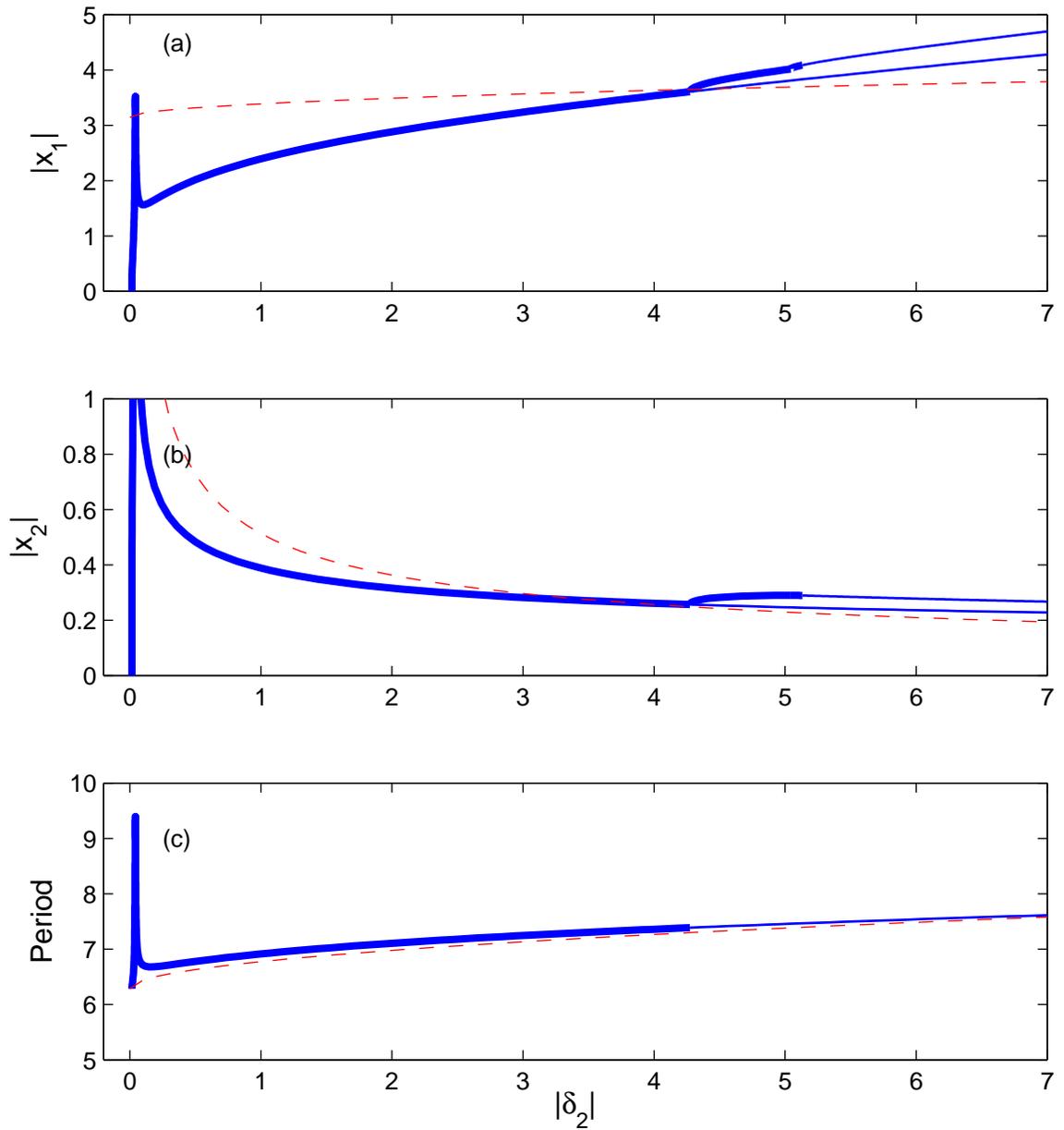}
\caption{Comparison of numerical bifurcation results (solid) to the analytical results
(dashed) from the map in Sec.~\ref{s:strong}, specifically, Eqs.~(\ref{e:fixpoints}).
Parameters are the same as in Fig.~\ref{f:bifdiafull}. }
\label{f:pulseharm_map}
\end{figure}

For ``large" values of the coupling, when $\delta_2 = O(1)$, the intensity of
laser-1 becomes pulsating, while the oscillations of laser-2 remain small
and nearly harmonic (see Fig.~\ref{f:timeplot_bpd}). We derive an iterated
map to describe the oscillations when $\delta_2 = O(1)$. Fixed points of the
resulting map correspond to periodic solutions of Eq.~(\ref{e:laserxy}). Our results
are summarized in Fig.~\ref{f:pulseharm_map}, where we compare the amplitudes
and period to those obtained from numerical simulation.

To construct the map we take advantage of the fact that the intensity of laser-1 has
two distinct regimes: during the pulse when $y_1 \gg 1$, and a long interval of time
when $y_1 \approx -1$. For a single pulsating laser, the solutions to
Eq.~(\ref{e:laserxy}) have described using an iterated map constructed with the
method of matched asymptotics \cite{ScEr94,CaBiScTr00}; we will use the same
approach here and so will just summarize our results.
We will first find the ``outer" solutions to
Eqs.~(\ref{e:laserxy}) with the approximation $y_1 \approx -1$ (
see Fig.~\ref{f:timeplot_bpd}a from $t_0$ to $t_1$). We will then
reanalyze the coupled system with an ``inner" or ``boundary-layer" approximation
$y_1 \gg 1$ (see Fig.~\ref{f:timeplot_bpd}a from $t_1$ to $t_2$). The typical next
step is to match the inner and outer solutions
to form a composite solution over the whole period. However, we are interested
in the dynamics from one pulse to the next. Hence, we will simply patch the
solutions together to form an iterated map.

As described above, the pulsations of laser-1 define the inner and outer regimes.
However, laser-2 continues to exhibit small-amplitude, nearly-harmonic
oscillations. Thus, in each regime we will use the method of multiple scales to
describe the oscillations of laser-2.

\subsection{Re-supply of the inversion, $y_1\approx -1$}

We first consider the outer regime when $y_1 \approx -1$. We define $t=t_0$ as the time of
completion of a previous pulse, when $y_1 = 0$ and the inversion $x_1$ is at its minimum
(see Fig.~\ref{f:timeplot_bpd}). The end of the outer regime will be defined to be when
the intensity increases from $y_1 = -1$ back to $y_1=0$ and the inversion $x_1$ is at
its maximum.

We first consider laser-2. When $y_1\approx -1$ the dynamics of laser-2 can be approximated as
\begin{eqnarray}
	\frac{dy_{2}}{dt} & = & \beta x_{2}(1+y_{2}),\nonumber\\
	\frac{dx_{2}}{dt} & = & \beta[-y_{2}-\epsilon\beta x_{2}(a_{2}+by_{2})-\delta_{1}],
	\label{e:laserxy2outer}
\end{eqnarray}
We can solve this system using the method of multiple scales as we did in
Sec.~\ref{s:two} under the assumption that $\delta_1$ is small ($\delta_1=O(\epsilon)$).
We find that to leading order laser-2 is a weakly damped, nearly harmonic oscillator
described as
\begin{eqnarray}
   x_2(t) & = & e^{-\frac{1}{2}\epsilon a_2 \phi},
     [ x_{20} \cos(\omega \phi) - y_{20} \sin(\omega \phi) ],\;\;\;\phi= t-t_0,\nonumber\\
   y_2(t) & = & -\frac{dx_2}{dt},\nonumber\\
   \omega & = &  1 -\frac{1}{2}\delta_1 - \frac{1}{24}(x_{20}^2+y_{20}^2)e^{-\epsilon a_2 \phi},
   \label{e:l2outer}
\end{eqnarray}
where $(x_2(t_0),y_2(t_0))=(x_{20},y_{20})$ is the state of laser-2 at the end
of the previous pulse. (As in Sec.~\ref{s:two}, $x_2$ and $y_2$ are $O(\epsilon^{1/2})$
such that the next term in the solutions in Eqs.~(\ref{e:l2outer}) would be
$O(\epsilon)$.)

We now examine laser-1 in more detail. With $y_1 \approx -1$ we have
\begin{equation}
  \frac{dx_1}{dt} = 1 - \epsilon (a_1-b) x_1 + \delta_2 y_2,
\end{equation}
which can be integrated to obtain
\begin{equation}
  x_1(t) = (x_{10} - \frac{1}{\gamma})e^{-\gamma \phi} + \frac{1}{\gamma}
      + \delta_2 e^{-\gamma t}\int_{t_0}^t e^{\gamma s} y_2(s) ds,
\end{equation}
where $\gamma = \epsilon (a_1-b)$.
Using the result for $y_2$ from Eqs.~\ref{e:l2outer} we obtain
\begin{eqnarray}
  x_1(t) &=& (x_{10} - \frac{1}{\gamma})e^{-\gamma \phi} + \frac{1}{\gamma} \nonumber\\
   &&+ \frac{\delta_2}{\alpha^2+\omega^2} e^{-\gamma \phi}\left[
        (\alpha y_{20}-\omega x_{20}) (e^{\alpha \phi}\cos(\omega \phi) -1)\right. \nonumber\\
   &&\quad\quad\quad\quad\quad\quad\quad \left.  + (\omega y_{20}+\alpha x_{20})e^{\alpha \phi}\sin(\omega \phi) \right],
	  \label{e:x1outer}
\end{eqnarray}
where $\alpha = (\gamma - \epsilon a_2/2)$. We can now use $x_1$ to improve our approximation
for $y_1$ by substituting Eq.~(\ref{e:x1outer}) in the equation for $y_1$ in Eq.~(\ref{e:laserxy});
we then integrate to give
\begin{equation}
  y_1(t) = -1 + (1+y_{10}) \exp\left( \int_{t_0}^t x_1(s)ds \right).
  \label{e:y1outer}
\end{equation}

During the outer regime the inversion grows almost linearly from its minimum
to maximum values, more precisely, for $\epsilon \ll 1$, $x_1 \approx (t-t_0)$. With the
inversion re-supplied the laser can then emit a new pulse of light. We define
the start of the next pulse at $t=t_1$ to be when $y_1(t_1) = 0$. Thus, the next
pulse begins when the integral in the exponential of Eq.~(\ref{e:y1outer}) is zero, or
\begin{equation}
  \int_{t_0}^{t_1} x_1(t)dt = 0.
  \label{e:Pouter}
\end{equation}

\subsection{Pulse regime, $y_1 \gg 1$}

The inner regime is defined to be when the intensity is large, $y_1 \gg 1$, and occurs
over a very short interval of time.  Specifically, if $y_1 = O(E)$, where $E\gg 1$ is
related to the energy, then $x_1 = O(E^{1/2})$ and the width of the pulse is
$O(1/E^{1/2})$ \cite{CaBiScTr00}. On the other hand, the oscillations of laser-2
remain small, $O(\epsilon^{1/2})$. Thus, we assume that to leading order,
laser-2 has no effect on laser-1 during the pulse. We can then use the results
from \cite{CaBiScTr00}, in the absence of modulation, to describe laser-1.
Namely, (i) the end of the pulse, $t=t_2$, is defined to be when the pulse
intensity returns to zero,
$y(t_2)= 0$, (ii) the width of the pulse is negligible compared to the time in the
outer regime, $(t_2 - t_1)\approx 0$, and (iii) the inversion drops from its
maximum to minimum value with reduction due to damping:
\begin{equation}
   x_1(t_2) = -x_1(t_1) + \frac{2}{3} \epsilon b x_1(t_1)^2.
   \label{e:x1inner}
\end{equation}
($x_1(t_2)$ is negative at the minimum so that the additional positive term
is a reduction in the magnitude of the minimum.)

The large and narrow ($t=O(1/E^{1/2})$) pulse of laser-1 does have
a significant effect on laser-2. To determine the appropriate inner problem
for laser-2, we scale the pulse amplitude as $y_1=O(E)$ and stretch time
according to $t=O(1/E^{1/2})$. The coupling is weak with $\delta_1 = O(\epsilon)$.
Finally, we assume that $E = O(1/\epsilon^{1/2})$ to obtain
\begin{eqnarray}
  \frac{d x_2}{dt} &=& \delta_1 y_1,\nonumber\\
  \frac{d y_2}{dt} &=& 0.
\end{eqnarray}
Thus, to leading order $y_2$ is constant during the pulse while $x_2$ is given by
\begin{equation}
   x_2(t_2) - x_2(t_1)= \delta_1 \int_{t_1}^{t_2} y_1(t) dt.
\end{equation}
However, in the pulsing regime we have that $dx_1/dt \approx -y_1$ so that
\begin{equation}
   x_2(t_2) - x_2(t_1) = -\delta_1 \int_{t_1}^{t_2} \frac{dx_1}{dt} dt
                      = \delta_1 [x_1(t_1)-x_1(t_2)].
   \label{e:x2inner}
\end{equation}
Thus, for laser-2 we have that
\begin{equation}
   x_2(t_2) = x_2(t_1) + \delta_1 2 x(t_1),\quad y(t_2)=y(t_1)
   \label{e:xy2inner}
\end{equation}
where we have used Eq.~(\ref{e:x1inner}) in Eq.~(\ref{e:x2inner}) and kept only
the leading order terms (the leading order is $O(\epsilon^{1/2})$ and we have
dropped the $O(\epsilon)$ corrections). The net effect is that at the
end of the pulse when $t=t_2$, the intensity $y_2$ remains unchanged, while
$x_2$ has received a ``kick" due to the pulse from laser-1.

\subsection{Constructing the map}

To construct a map, we ``patch" together the results from the outer and inner analysis
of the previous two sections.

For laser-2 we initially have $(x_2(t_0),y_2(t_0))=(x_{20},y_{20})$ that in the
outer region evolves according to Eqs.~(\ref{e:l2outer}) until $t=t_1$. Then, in the
inner region, laser-2 receives the pulse from laser-1 according to Eq.~\ref{e:xy2inner}.
Thus, we have that
\begin{equation}
\left[ (x_2(t_0),y_2(t_0))=(x_{20},y_{20}) \right] \mapsto
(x_2(t_1),y_2(t_1)) \mapsto (x_2(t_2),y_2(t_2)).
\end{equation}
The total time from one pulse to the next is $t_2-t_0$. However, because the pulse
is so short ($O(\epsilon^{1/2})$), we make the approximation that $t_2\approx t_1$ and
define the total time as $P= t_1-t_0$. Finally, for notation convenience we
define the intermediate value of the inversion of laser-1 as $G(P)=x_1(t_1)$.
The map for laser-2 is then
\begin{eqnarray}
   x_{2}  &\mapsto &  e^{-\frac{1}{2}\epsilon a_2 P}
     [ x_{2} \cos(\omega P) - y_{2} \sin(\omega P) ]
     + \delta_1 2 G(P) ,\nonumber\\
   y_{2} & \mapsto & e^{-\frac{1}{2}\epsilon a_2 P}
     [ x_{2} \sin(\omega P) + y_{2} \cos(\omega P) ],
     \label{e:x2y2map}
\end{eqnarray}
where
\begin{eqnarray}
  G(P) &=& (x_{1} - \frac{1}{\gamma})e^{-\gamma P} + \frac{1}{\gamma} \nonumber\\
   &&+ \frac{\delta_2}{\alpha^2+\omega^2} e^{-\gamma P}\left[
        (\alpha y_{2}-\omega x_{2}) (e^{\alpha P}\cos(\omega P) -1)
	  + (\omega y_{2}+\alpha x_{2})e^{\alpha P}\sin(\omega P) \right].
  \label{e:Gmap}
\end{eqnarray}

The time from one pulse to the next is determined when $y_1 =0$ with $t_2\approx t_1$.
Thus, from Eq.~(\ref{e:Pouter}) we have a condition to determine $P$ as
\begin{equation}
  \int_0^P G(t) dt = 0.
  \label{e:Pmap}
\end{equation}

Finally, for the inversion of laser-1 we have
\begin{equation}
	x_1(t_0) \mapsto [x_1(t_1) = G(P)] \mapsto x_1(t_2),
\end{equation}
yielding
\begin{equation}
   x_{1} \mapsto -G(P) + \frac{2}{3} \epsilon b G(P)^2.
   \label{e:x1map}
\end{equation}

The map is evaluated as follows: (i) The current state of the system,
given by $x_1$, $x_2$ and $y_2$, is known. (ii) Compute the time $P$
of the next pulse using Eq.~(\ref{e:Pmap}). (iii) With $P$ fixed
we can evaluate $G(P)$ in Eq.~(\ref{e:Gmap}). (iv) The current state
of the system and $G(P)$ determine new values
for $x_2$ and $y_2$ with Eqs.~(\ref{e:x2y2map}). (v) Finally,
$x_1$ is found from Eq.~(\ref{e:x1map}). Summarizing, we have
\begin{eqnarray}
   \mbox{(ii) Eq.~(\ref{e:Pmap})} &\quad & f_1 (P;x_1,x_2,y_2)=0,\nonumber\\
   \mbox{(iv) Eq.~(\ref{e:x2y2map})}&\quad & x_2 \mapsto f_2(x_2,y_2,G(P))\nonumber\\
                  &\quad & y_2 \mapsto f_3(x_2,y_2,G(P)) \nonumber\\
   \mbox{(v) Eq.~(\ref{e:x1map})}& \quad & x_1 \mapsto f_4(G(P)).
   \label{e:mapsummary}
\end{eqnarray}

\subsection{Periodic solutions as fixed points}

Fixed points of the map described by Eqs.~(\ref{e:mapsummary}) correspond to
periodic solutions of the original flow, Eqs.~(\ref{e:laserxy}). However, it
is not feasible to analyze the map without further approximations.
We will look for fixed points making use of $\epsilon \ll 1$. With
heavy use of symbolic computation, we find that the maximum amplitudes and the
period of the oscillations are given by
\begin{eqnarray}
	\max[ x_1] &=& \pi + \sqrt{ \frac{3a_2}{2a_1}\delta_1 |\delta_2|},\nonumber\\
        \max[x_2] &=& \max [y_2] = \sqrt{ \pi^2\frac{2a_1}{3a_2} \frac{\delta_1}{|\delta_2|}},
	\nonumber\\
        P &=& 2\max[x_1].
	\label{e:fixpoints}
\end{eqnarray}
For each result the neglected terms are $O(\epsilon)$.
In addition, we also obtain the phase relationship result that $x_2 \approx 0$ when
$x_1$ is at its minimum, which is consistent with Fig.~\ref{f:timeplot_bpd}. We have
plotted the predictions of Eqs.~(\ref{e:fixpoints}) along with the results from
numerical simulations
in Fig.~\ref{f:pulseharm_map} and they show good agreement, where for clarity we have
removed the higher bifurcation branches present in Fig.~\ref{f:bifdiafull}.

The period $P$ and the amplitude of laser-2 show excellent agreement. We see that
$\max[ x_2 ] \approx 1/|\delta_2|^{1/2}$. This may initially seem counter intuitive
because the pulses of laser-1 grow with $|\delta_2|$ and provide a greater kick to
laser-2. Indeed, a leading-order
approximation to the kick applied by laser-1
$\delta_1 2 G(P) \approx \delta_1 P$, thus, the strength of the kick increases as
the period increases. However, with longer periods the exponential decay due to
damping in the outer regime has more time to decrease the amplitude of laser-2.
The net effect is a decrease in
$\max[ x_2]$ with increasing $|\delta_1|$.

The net coupling strength of laser-2 to laser-1, with respect to $|\delta_2|$,
is $\delta_2 y_2 = O(\delta_2^{1/2})$. Thus,
the amplitude of the pulsations
increases with increasing $\delta_2$. The fit between the analysis and numerics
is not as good for laser-1. To achieve a better fit we need to derive a map
that includes higher-order terms in $\epsilon$, which we have not attempted.

\section{Discussion}
\label{e:discuss}

For two coupled lasers we have studied the bifurcations that occur when
the coupling constants are of opposite sign and unequal. Specifically,
the coupling is asymmetric in that we fix one coupling constant ($\delta_1>0$) to be
small, while varying the other ($\delta_2<0$). There are two Hopf bifurcations
to periodic output, one for $\delta_2$ positive and one for $\delta_2$ negative.
We have focused our attention on the latter because of its similarity to our
work with delay coupling in \cite{KiRoArCaSc05}. When the output of laser-2 is
nearly harmonic, the negative coupling effectively corresponds to phase shift by
half of a period. This is equivalent to delayed coupling when the delay is
half the period.

As $|\delta_2|$ is increased there is an initial Hopf
bifurcation from the laser's CW steady-state to periodic solutions.
We then observe two resonance regimes where the coupled system shows
novel and interesting output. (i) Close to the Hopf
bifurcation a resonance can occur where the amplitude of the laser oscillations
becomes large. This is unexpected because both coupling constants are still small.
The resonance is due to the negative-coupling that effectively reduces the
damping in the laser. (ii) As the strength of the coupling
increases further the periodic solutions may, depending on the parameters,
exhibit a period-doubling sequence to chaos as well as the coexistence of
subharmonic solutions. These effects are reminiscent of a periodically
modulated laser. In the case of the coupled lasers the laser receiving
the weak coupling remains a nearly harmonic oscillator that excites the
strong resonances of the pulsating laser.

The large-amplitude resonance that occurs for small coupling can be
easily understood from the well-known coupled Stuart-Landau equations given
generically by Eq.~(\ref{e:stuartlandau}). Steady-state solutions of
Eq.~(\ref{e:stuartlandau}) correspond to the amplitude of the periodic
solutions. Simple algebra shows that tuning some of the coupling parameters to
be of opposite sign can lead to a vanishing denominator.
Physically, the coupling term is providing an effective negative-damping
that cancels with the lasers self-damping and, hence, a resonance effect.
Our analysis assumed that both coupling constants were of the same relative
size, $\delta_j = O(\epsilon)$. However, other scalings satisfy the Hopf condition,
e.g., $\delta_1 = O(\epsilon^{1/2})$ and $\delta_2 = O(\epsilon^{3/2}$. This does
not change the qualitative properties of the bifurcating periodic solutions in
any way.

In App.~\ref{s:pulsations} we have attempted to describe the solutions
that occur near the peak of the resonance. In this regime, both lasers show
approximately equal amplitude pulsating solutions as exhibited by
Fig.~\ref{f:reson}(a1). Our analysis reproduces the
equation for the location of the resonance $\delta_{2S}$. It also
predicts that the period should be twice the maximum amplitude of the
inversion. That both the period and the amplitude show a resonance peak
at $\delta_{2S}$ in Fig.~\ref{f:reson}a \& b is consistent with this
result but the scale factor of 2 is not correct. Also, we do not obtain
and expression for how the period (or amplitude) depends on the parameter
$\delta_2$. A difficult higher order analysis would be required to remedy
these last two limitations.

In Fig.~\ref{f:timeplot_bpd} we showed that when $|\delta_2| = O(1)$
(see Fig.~\ref{f:bifdiafull}) that one of the lasers amplitudes is
large while the other's remains small. The term ``localized solutions" has
been used to describe the case when \textit{identical} oscillators in a
coupled system exhibit amplitudes of different scales. In coupled lasers
localized solutions have been described by Kuske and Erneux \cite{KuEr97}
who derived a similar pair of integral conditions to Eq.~(\ref{e:submel}).
Instead of looking for pulsating solutions, they considered $O(1)$ solutions
approximated using a Poincare-Lindstedt method, and small amplitude solutions
approximated with the method of multiple scales. Repeating this analysis
for our problem reproduces the Hopf bifurcation results that we obtained in
Sec.~\ref{s:harmonic}.

To describe the system's output when the coupling is strong, we have derived
a map that predicts the period, amplitude and phase of the lasers from one
pulse of laser-1 to the next. Constructing the map relies on combining
both strongly and weakly nonlinear asymptotic methods. That is, we used
matched asymptotics to describe the pulsating laser-1 and to separate one
period into an inner and outer subintervals. For the small-amplitude laser-2
we used the multiple scale methods within each subinterval. We obtain
very good agreement between the amplitude of laser-2 and the overall period.
The amplitude of the pulsations of laser-1 are not described quite as well.
This could be because we need to consider higher-order terms in our solutions,
or we are comparing our numerical and analytical results in a less than
ideal parameter regime.

The large-amplitude solutions in the resonance peak just after the Hopf bifurcation
are not the same as those that appear due to a ``singular-Hopf
bifurcation" \cite{BaEr86,BaEr92}. In the latter case, the large amplitude
oscillations are due to crossing a separatrix separating small-amplitude
solutions near the Hopf bifurcation from large-amplitude relaxation oscillations
formed around a slow manifold. The functional form of the dissipation terms
in the present problem disallows this type of behavior.

To our knowledge, the small-coupling resonance peak has not been previously described;
most likely this due to consideration of physical systems where controlling
the sign of the coupling is not possible. However, in a forthcoming study
we will show that for same-sign coupling but with delay, we can again
produce the resonance because the delay provides the phase shift that effectively
leads to the sign change.

\appendix

\section{Pulsating solutions for weak coupling}
\label{s:pulsations}

In Sec.~\ref{s:harmonic} we looked for small-amplitude solutions
near the Hopf bifurcation point. We now allow the amplitude of the solution
to be arbitrary but will still consider the coupling to be small. The laser
system Eq.~(\ref{e:laserxy}) can be rewritten so that the intensity and
inversion evolve according to a perturbed-Hamiltonian system \cite{CaBiScTr00}.
From the coupled-Hamiltonian systems, we derive solvability conditions for
T-periodic solutions as
\begin{equation}
	\int_0^{T} (- a_j x_j^2 + d_k x_j y_k) dt = 0,
	\label{e:submel}
\end{equation}
The integrals in Eq.~(\ref{e:submel}) are computed by evaluating $x_j$ and $y_j$
on periodic orbits of the $\epsilon=0$, Hamiltonian system. Unfortunately, we do not
have closed form analytical solutions for $x_j$ and $y_j$. However, for pulsating
output we can construct approximate solutions to the Hamiltonian system
using matched asymptotic expansions similar to what we did in Sec.~(\ref{s:strong}).
In this case, we match the outer and inter solutions to determine a uniform
solution that can be used to evaluate the integrals in Eq.~(\ref{e:submel}).
Because we have carried out similar
calculations in the past \cite{ScEr94,CaBiScTr00},
we will only summarize the details of the intermediate steps.

Before proceeding, we mention that Kuske and Erneux \cite{KuEr97} derived an almost equivalent
pair of solvability conditions for two coupled lasers. Their goal was to
investigate so-called ``localized" solutions where one laser has $O(1)$ amplitude
oscillations, while the other has small oscillations, and both are approximated
using the Poincare'-Linstedt perturbation method. Doing this calculation for
our problem effectively reproduces our earlier results obtained near the
Hopf bifurcation point and thus does not provide new information.

We assume that laser-$j$ has period $T_j$, which is some fraction of
the total period with $T = n_j T_j$. The first term in each integral is
$x_j^2$ and because it does not involve the other laser is independent
of the phase relationship between lasers $j$ and $k$. Thus, using the
results from
\cite{CaBiScTr00} we have
\begin{equation}
  \int_0^{T} x_j^2 dt = \frac{n_j}{12}T_j^3.
\end{equation}

Integrating $x_j y_k$ is more complicated because we must allow for a
phase difference, $T_{\phi}$, between the two pulsating lasers. However, it is easy to
predict the form of the result. The intensity $y_k$ is pulsating and acts like
a delta function that samples the inversion $x_j$ at the time of the pulse.
The effect of the integral is to sum all of the sample values of the population
inversion. In effect, we have a pulse train due to one laser sampling
the population inversion of the other.  After carrying out the detailed calculations
based on the approximate solutions of the Hamiltonian system, we obtain our final
result for both solvability conditions:
\begin{equation}
	-a_{1}\frac{n_1}{12}T_1^3 + T_2 \delta_2 \sum_{k=0}^{n_2-1}
	x_1((k+\frac{1}{2})T_2)=0,
\end{equation}
where
\begin{equation}
	x_1(t) = \left\{ \begin{array}{cc}-jT_1 + T_{\phi} + t,
	&  jT_1 -T_{\phi}< t\le (j+\frac{1}{2})T_1 - T_{\phi}\\
        -(j+1)T_1 + T_{\phi} + t,&
         (j+\frac{1}{2})T_1-T_{\phi} < t\le (j+1)T_1-T_{\phi}
        \end{array}\right. ,
\end{equation}
and
\begin{equation}
  -a_{2}\frac{n_2}{12}T_2^3 + T_1 \delta_1 \sum_{j=0}^{n_1-1}
   x_2((j+\frac{1}{2})T_1-T_{\phi})=0,
\end{equation}
where
\begin{equation}
  x_2(t) = \left\{ \begin{array}{cc}
        -kT_2 + t,&  kT_2 < t\le (k+\frac{1}{2})T_2\\
        -(k+1)T_2 + t,& (k+\frac{1}{2})T_2 < t\le (k+1)T_2
        \end{array}\right. .
\end{equation}
The inversion variable $x_j$ of each laser is a saw-toothed type function that
increases linearly from the time of the previous pulse to the next. Specifically,
for laser-2, $x_2$ increases from 0 (at time $k T_2$) to $x_2 = T_2/2$. The intensity
pulse depletes the inversion to $x_2=-T_2/2$, whereupon $x_2$ then increases
linearly back to 0. Laser-1 is the same except that we must allow for a phase time
$T_{\phi}$ between the two lasers.

We consider the simple case of a 1:1 resonance between the lasers where
$T = T_1 = T_2$ so that $n_1= n_2 = 1$. The solvability conditions reduce to
\begin{equation}
  -\frac{a_1}{12}T^2 + \delta_2 x_1(\frac{T}{2})=0
\end{equation}
\begin{equation}
  -\frac{a_2}{12}T^2 + \delta_1 x_2(\frac{T}{2}-T_{\phi})=0
\end{equation}
Because we are interested in periodic solutions, it is sufficient to consider
$0\le T_{\phi} \le T$. Then, substituting for $x_1$ and $x_2$, the conditions reduce to
\begin{equation}
  -\frac{a_{1}}{12}T^2 + \delta_2 (-\frac{T}{2} + T_{\phi})=0
\end{equation}
\begin{equation}
  -\frac{a_{2}}{12}T^2 + \delta_1 (\frac{T}{2} - T_{\phi})=0
  \label{e:rescond11}
\end{equation}
After eliminating the phase $T_{\phi}$, we obtain
\begin{equation}
  (1 + \frac{a_{2} \delta_2}{a_{1} \delta_1})T^2 = 0.
\end{equation}
For periodic solutions with $T \ne 0$, we are forced to set the term in
parenthesis equal to zero. This is exactly the same condition that identifies
the location of the singularity in the Hopf bifurcation equation (\ref{e:delta2s}).
This confirms that there is an equal-amplitude, pulsating 1:1 resonance between
the lasers when $\delta_2=\delta_{2S}$. However, we do not have any information
on the period or amplitude, which would require continuing the analysis to higher order.



\end{document}